\documentclass[aps,prl,groupedaddress]{revtex4}
\usepackage{epsfig}

\begin{document}

\title{Fair Solution to the Reader-Writer-Problem with Semaphores only}

\author{H. Ballhausen}
\email{physics@ballhausen.com}
\affiliation{Institute for Theoretical Physics, Heidelberg University\\Philosophenweg 16, 69120 Heidelberg, Germany}

\begin{abstract}
\noindent
The reader-writer-problem is a standard problem in concurrent programming. A resource is shared by several processes
which need either inclusive reading or exclusive writing access. The known solutions to this problem typically involve 
a number of global counters and queues. Here a very simple algorithm is presented which needs only two semaphores for
synchronisation and no other global objects. The approach yields a fair solution without starving.
\end{abstract}

\maketitle

\subsection{The reader-writer-problem}

\noindent
Assume the following situation: a shared resource, say a database, is accessed by several concurrent processes for reading or 
writing. Because the readers do not change the database, any number of them may be granted simultaneous access. On the other 
hand, only one writer is allowed to modify the database at a time.

\subsection{The standard solution}

\noindent
The standard solution to the reader-writer-problem using semaphores as found in standard literature
\cite{Lit1} and countless lecture scripts reads as follows ( in self explanatory pseudo code ):

\bigskip
\noindent
\begin{tabular}{|lllll|}
\hline
$~~~$ & $~~~$ & $~~~$ & $~~~$ & $~~~$ \\
$~~~$ & {\tt {\bf var} integer num = 0;																} & $~~~$ & {\tt																													} & $~~~$ \\
$~~~$ & {\tt {\bf var} semaphore mutex = 1;														} & $~~~$ & {\tt																													} & $~~~$ \\
$~~~$ & {\tt {\bf var} semaphore access = 1;													} & $~~~$ & {\tt																													} & $~~~$ \\
$~~~$ & {\tt																													} & $~~~$ & {\tt																													} & $~~~$ \\
$~~~$ & {\tt {\bf process} reader ( integer i = 1 ... m ) \{					} & $~~~$ & {\tt {\bf process} writer ( integer j = 1 ... n ) \{					} & $~~~$ \\
$~~~$ & {\tt $~~~$ {\bf do} \{																				} & $~~~$ & {\tt $~~~$ {\bf do} \{																				} & $~~~$ \\
$~~~$ & {\tt $~~~~~~$ {\bf P}(mutex);																	} & $~~~$ & {\tt $~~~~~~$ {\bf P}(access);																} & $~~~$ \\
$~~~$ & {\tt $~~~~~~$ num=num+1;																			} & $~~~$ & {\tt $~~~~~~$ // ... writing ...															} & $~~~$ \\
$~~~$ & {\tt $~~~~~~$ {\bf if}(num==1) {\bf P}(access);								} & $~~~$ & {\tt $~~~~~~$ {\bf V}(access);																} & $~~~$ \\
$~~~$ & {\tt $~~~~~~$ {\bf V}(mutex);																	} & $~~~$ & {\tt $~~~~~~$ // ... other operations ...											} & $~~~$ \\
$~~~$ & {\tt $~~~~~~$ // ... reading ...															} & $~~~$ & {\tt $~~~$ \}																									} & $~~~$ \\
$~~~$ & {\tt $~~~~~~$ {\bf P}(mutex);																	} & $~~~$ & {\tt \}																												} & $~~~$ \\
$~~~$ & {\tt $~~~~~~$ num=num-1;																			} & $~~~$ & {\tt																													} & $~~~$ \\
$~~~$ & {\tt $~~~~~~$ {\bf if}(num==0) {\bf V}(access);								} & $~~~$ & {\tt																													} & $~~~$ \\
$~~~$ & {\tt $~~~~~~$ {\bf V}(mutex);																	} & $~~~$ & {\tt																													} & $~~~$ \\
$~~~$ & {\tt $~~~~~~$ // ... other operations ...											} & $~~~$ & {\tt																													} & $~~~$ \\
$~~~$ & {\tt $~~~$ \}																									} & $~~~$ & {\tt																													} & $~~~$ \\
$~~~$ & {\tt \}																												} & $~~~$ & {\tt																													} & $~~~$ \\
$~~~$ & $~~~$ & $~~~$ & $~~~$ & $~~~$ \\
\hline
\end{tabular}

\bigskip
\noindent
The above algorithm uses a global variable {\tt num} to keep track of the number of readers currently
reading the database. The first reader to enter and the last reader to leave requests respectively
releases the database through the binary semaphore {\tt access}. The same semaphore is used by the
writers to ensure exclusive access. The binary semaphore {\tt mutex} finally protects the critical
sections involving the global counter. 

\bigskip
\noindent
The solution has two drawbacks, though. On the one hand, readers may enter and leave the database all 
the time without ever releasing waiting writers. This reader preference renders the solution unfair
and leads to starving. On the other hand, global data is often inefficient as compared to pure 
synchronization objects like semaphores. And extensions of the algorithm above making it fair
need even more global data like queues etc.

\newpage
\subsection{A fair solution}

\noindent
In fact, there is a very simple algorithm using only semaphores still offering a fair solution:

\bigskip
\noindent
\begin{tabular}{|lllll|}
\hline
$~~~$ & $~~~$ & $~~~$ & $~~~$ & $~~~$ \\
$~~~$ & {\tt {\bf var} semaphore mutex = 1;													} & $~~~$ & {\tt																											}	& $~~~$ \\
$~~~$ & {\tt {\bf var} semaphore access = m;												} & $~~~$ & {\tt																											} & $~~~$ \\
$~~~$ & {\tt 																												} & $~~~$ & {\tt																											} & $~~~$ \\
$~~~$ & {\tt {\bf process} reader ( integer i = 1 ... m ) \{				} & $~~~$ & {\tt {\bf process} writer ( integer j = 1 ... n ) \{			} & $~~~$ \\
$~~~$ & {\tt $~~~$ {\bf do} \{																			} & $~~~$ & {\tt $~~~$ {\bf do} \{																		} & $~~~$ \\
$~~~$ & {\tt $~~~~~~$ {\bf P}(access);															} & $~~~$ & {\tt $~~~~~~$ {\bf P}(mutex);															} & $~~~$ \\
$~~~$ & {\tt $~~~~~~$ // ... reading ...														} & $~~~$ & {\tt $~~~~~~$ for k = 1 ... m do {\bf P}(access);					} & $~~~$ \\
$~~~$ & {\tt $~~~~~~$ {\bf V}(access);															} & $~~~$ & {\tt $~~~~~~$ // ... writing ...													} & $~~~$ \\
$~~~$ & {\tt $~~~~~~$ // ... other operations ...										} & $~~~$ & {\tt $~~~~~~$ for k = 1 ... m do {\bf V}(access);					} & $~~~$ \\
$~~~$ & {\tt $~~~$ \}																								} & $~~~$ & {\tt $~~~~~~$ {\bf V}(mutex);															} & $~~~$ \\
$~~~$ & {\tt \}																											} & $~~~$ & {\tt $~~~~~~$ // ... other operations ...			 						} & $~~~$ \\
$~~~$ & {\tt 																												} & $~~~$ & {\tt $~~~$ \}			 																				} & $~~~$ \\
$~~~$ & {\tt 																												} & $~~~$ & {\tt \}			 																							} & $~~~$ \\
$~~~$ & $~~~$ & $~~~$ & $~~~$ & $~~~$ \\
\hline
\end{tabular}

\bigskip
\noindent
The idea is here that one reader alone takes up the same space as all the readers together.
This is realized with a generalized semaphore {\tt access}. At the beginning its value is
equal to the total number of readers. Everytime a reader enters the database it is 
decremented and everytime a reader leaves it is incremented. That way the readers alone
are always granted simultaneous access. 

\bigskip
\noindent
But when a writer wants to enter the database it occupies all space step by step waiting 
for old readers to leave while denying entrance to new ones. The binary semaphore {\tt mutex} 
is needed to prevent deadlocks when say two entering writers occupy half of the space each.

\subsection{Conclusion}

\noindent
An algorithm for the reader-writer-problem has been presented which uses only semaphores for
synchronization and no other global data. It offers a fair solution without reader-preference,
writer-preference or starving.

\end{document}